\documentclass[twocolumn,superscriptaddress,aps,prb,longbibliography]{revtex4-2}
\usepackage{graphicx}
\usepackage{newtxtext,newtxmath,amsmath,mathtools}
\usepackage[colorlinks]{hyperref}

\begin{document}

\title{Microscopic model for the ground state, 1/3 plateau and excitations of $\gamma$-Mn$_3$(PO$_4$)$_2$} 

\author{P. A. Maksimov} 
\email{maksimov@theor.jinr.ru}
\affiliation{Bogolyubov Laboratory of Theoretical Physics, Joint Institute for Nuclear Research, Dubna, Moscow region 141980, Russia}
\author{L. V. Shvanskaya}
\affiliation{Geology Faculty, Moscow State University, Moscow 119991, Russia}
\author{O. S. Volkova}
\affiliation{Physics Faculty, Moscow State University, Moscow 119991, Russia}
\author{A. N. Vasiliev}
\affiliation{Physics Faculty, Moscow State University, Moscow 119991, Russia}

\begin{abstract}
We present a magnetic model for an antiferromagnetic compound $\gamma$-Mn$_3$(PO$_4$)$_2$, which was previously shown to exhibit a 1/3 magnetization plateau due to the trimer-based structure of the lattice of magnetic Mn$^{2+}$ ions with $S=5/2$. An exchange Hamiltonian that yields observed field transitions is obtained from fitting magnetization data.  It is shown that both biquadratic coupling and single-ion anisotropy are necessary to be present in the magnetic model to explain multiple phase transitions in the magnetic susceptibility data. The calculated magnetic spectrum is in agreement with the low-temperature specific heat data.
\end{abstract}

\maketitle


\section{Introduction}
\label{sec_intro}
Transition metal phosphates are known to exhibit intriguing magnetic properties at low temperatures due to an interplay of their peculiar crystal structure and anisotropic exchanges \cite{Shvanskaya_2020}.  One of these compounds, Mn$_3$(PO$_4$)$_2$, has been studied extensively and was shown to exhibit several polymorphs with varying magnetic behavior of its $S=5/2$ Mn$^{2+}$ ions \cite{Calvo_1969,Nord_1987,Bali_2000,Yan_2015}.  In particular, $\gamma$-Mn$_3$(PO$_4$)$_2$ \cite{Massa_2005}, an antiferromagnet with N\'eel temperature of $T_N=13.3$K, drew attention due to the presence of the 1/3 magnetization plateau \cite{Volkova_2016}. 

While magnetization plateaus have been studied as a signature of frustration and quantum effects of triangular lattice antiferromagnets \cite{chubukov91,Svistov_2003,expBaCoSbO2015}, they have also been found in a multitude of other circumstances \cite{vasiliev_plateaus}. One of the most notable examples is the Kitaev honeycomb antiferromagnet BaCo$_2$(AsO$_4$)$_2$ where the 1/3 plateau is stabilized by bond-dependent exchanges of cobalt ions, induced by sizeable spin-orbit coupling \cite{LP77,Cava_2020_BaCo,proximity_2023,Zapf_2024}. Moreover, even in materials with large spin, where quantum effects are small, magnetization plateaus can be stabilized by peculiar crystal structure \cite{SMPO_2008,SMPO_2009,SMPO_2014,vasiliev_2016} and exchanges beyond Heisenberg \cite{Svistov_2006,Ishii_2011,Zhou_2014,Jin_2024}.

It was previously argued that a trimer-based structure of the magnetic sublattice of $\gamma$-Mn$_3$(PO$_4$)$_2$ is responsible for the stabilization of the magnetization plateau \cite{Volkova_2016,Volkova_2025}. In this paper, we use previously published magnetization data to establish an effective magnetic Hamiltonian for $\gamma$-Mn$_3$(PO$_4$)$_2$. We show that the trimer-based crystal structure itself is not enough to exhibit the plateau, and either biquadratic coupling or single-ion anisotropy is necessary. In fact, Mn$^{2+}$ compounds are known to exhibit some form of anisotropy \cite{Edgar_1980,Hennion_2006,Pranzas_2006,Singleton_2016,Shatruk_2021}
or non-negligible biquadratic interactions \cite{Owen_1963}. Moreover, by carefully studying observed multiple magnetic phase transitions, we are able to extract parameters of the magnetic Hamiltonian and to show that \textit{both} biquadratic coupling and single-ion anisotropy are required to explain observed singularities in the magnetic susceptibility data. 

The paper is structured as follows. In Sec.~\ref{sec_ham}, we introduce the magnetic Hamiltonian and effective classical model. We outline details of fitting magnetization data in Sec.~\ref{sec_fit}. Various field-induced magnetic states are described in Sec.~\ref{sec_phases}. Magnetic excitations of the obtained model and its comparison to the experimental data are presented in Sec.~\ref{sec_spectrum}. We discuss our results in Sec.~\ref{sec_conclusion}.

\section{Magnetic Hamiltonian}
\label{sec_ham}
\begin{figure*}
    \includegraphics[width=\textwidth]{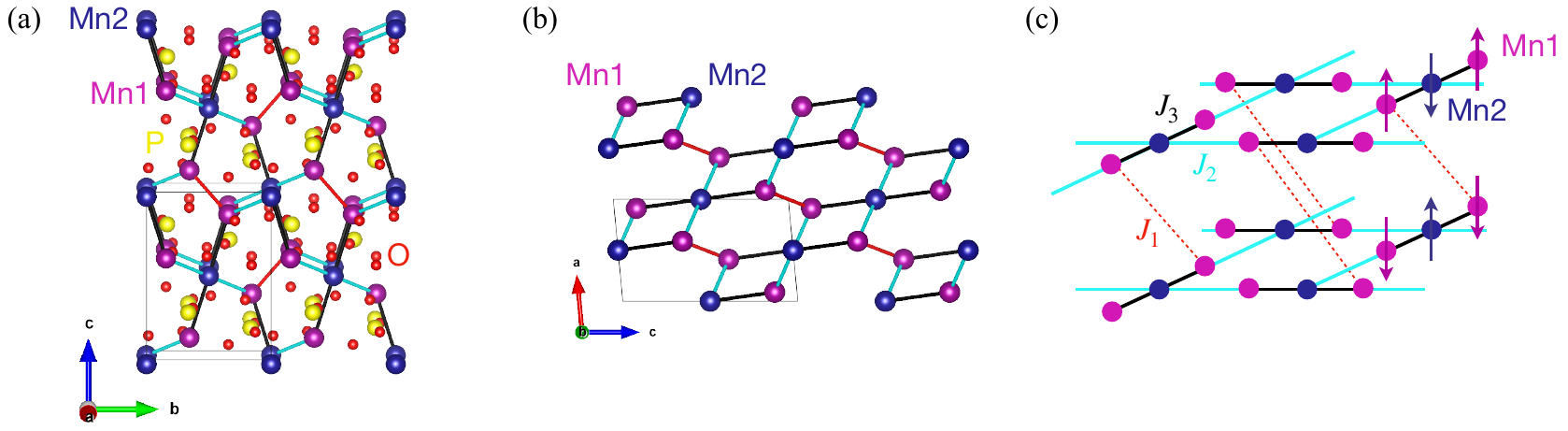}
\caption{(a) Crystal structure of $\gamma$-Mn$_3$(PO$_4$)$_2$. (b) Magnetic sublattice of $\gamma$-Mn$_3$(PO$_4$)$_2$ where only Mn atoms are shown. One can see that the lattice has a layered structure. (c) Schematic magnetic sublattice with exchange paths and magnetic ground state shown.}

    \label{fig:lattice}
\end{figure*}

In order to simulate magnetic properties of $\gamma$-Mn$_3$(PO$_4$)$_2$, we use the following Hamiltonian:

\begin{align}
    \mathcal{H}&=\sum_{\langle ij \rangle} J_{ij} \mathbf{S}_i \mathbf{S}_j-b_{ij}\left(\mathbf{S}_i \mathbf{S}_j \right)^2-\sum_i D_z \left( S^z_i\right)^2\nonumber\\&-g\mu_B \sum_i \mathbf{B}\cdot \mathbf{S}_i,
    \label{eq_ham}
\end{align}
where the first sum is over the bonds that are shown in Fig.~\ref{fig:lattice}(c) with three types of exchanges $J_{ij}$: $J_1$, $J_2$,  $J_3$. The geometry of the lattice can be described as layers of trimer-based decorated square lattice \cite{decorated_1951,decorated_2019,decorated_2020,Volkova_2025}, since it was shown using density-functional theory (DFT) calculations that $J_1 \ll J_2, J_3$ \cite{Volkova_2016}. 

Similarly, we introduce biquadratic exchanges on the same  bonds, $b_1 - b_3$. As we show below, both biquadratic exchange \cite{Mila_2000,Hotta_2018}, which are known to play a significant role in materials with large spin \cite{Walker_2020,Kartsev_2020,Chernyshev_EuC6}, and single-ion anisotropy $D_z$ are necessary to explain observed magnetic phases and transitions, including the 1/3 plateau. We assume $g=2$ \cite{Volkova_2016}.

Due to a rather large value of spin of Mn$^{2+}$ ions in $\gamma$-Mn$_3$(PO$_4$)$_2$, we approximate spins as classical vectors. While it is known that competition of several exchanges can lead to a complex magnetic configuration \cite{Sindzingre_2010,Gonzalez_2020,Mohylna_2022}, here, we assume four-sublattice magnetic states in the phase diagram: two spins of Mn1 and Mn2 sites on neighboring layers of the magnetic lattice of $\gamma$-Mn$_3$(PO$_4$)$_2$, as shown in Fig.~\ref{fig:lattice}(c).

Classical energy per site for such a four-sublattice state is given by
\begin{align}
E_\text{cl}/N=&\frac{1}{3} \left[J_1 \mathbf{S}_1 \mathbf{S}'_1+ \left(J_2+J_3\right) \left(\mathbf{S}_1 \mathbf{S}_2+\mathbf{S}'_1 \mathbf{S}'_2\right) \right]\nonumber\\
-&\frac{1}{3} \left\{ b_1  \left(\mathbf{S}_1 \mathbf{S}'_1\right)^2+ \left(b_2+b_3\right) \left[\left(\mathbf{S}_1 \mathbf{S}_2\right)^2+\left(\mathbf{S}'_1 \mathbf{S}'_2\right)^2\right] \right\} \nonumber\\
-&\frac{D_z}{6} \left[ 2 \left(S^z_1\right)^2+2 \left({S_1^z}'\right)^2+ \left(S^z_2\right)^2+ \left({S_2^z}'\right)^2\right]\nonumber\\&-\frac{1}{6} g\mu_B \mathbf{B}\cdot \left( 2 \mathbf{S}_1+2 \mathbf{S}'_1+\mathbf{S}_2+\mathbf{S}'_2\right),
\label{eq_ecl}
\end{align}
where $\mathbf{S}_1$, $\mathbf{S}_2$ are spins of Mn1 and Mn2 sites, and $\mathbf{S}'_1$, $\mathbf{S}'_2$ are the same on the neighboring layer. This expression can be minimized with respect to four spin directions to obtain ground state structure as a function of exchanges and magnetic field.

Exploring magnetic states in the relevant regime suggested by DFT values of exchanges \cite{Volkova_2016}:
\begin{align}
J_1=1.7\text{ K},~J_2=4.7\text{ K},~J_3=10.5\text{ K},
\label{eq_dft}
\end{align}
we found that only coplanar states are stabilized. Therefore, magnetic moments can be defined as  $\mathbf{S}_i=S\left(\cos \theta_i,0,\sin \theta_i \right)$ where $\theta_i$ is determined relative to the easy axis $z$. Numerical minimization of the energy \eqref{eq_ecl} over four variables $\theta_1,\theta_2,\theta'_1,\theta'_2$ yields magnetic ground state configurations, which we use to fit magnetization data from Ref.~\cite{Volkova_2016}.

\section{Tuning parameters}
\label{sec_fit}

\begin{figure*}
\centering
\includegraphics[width=\textwidth]{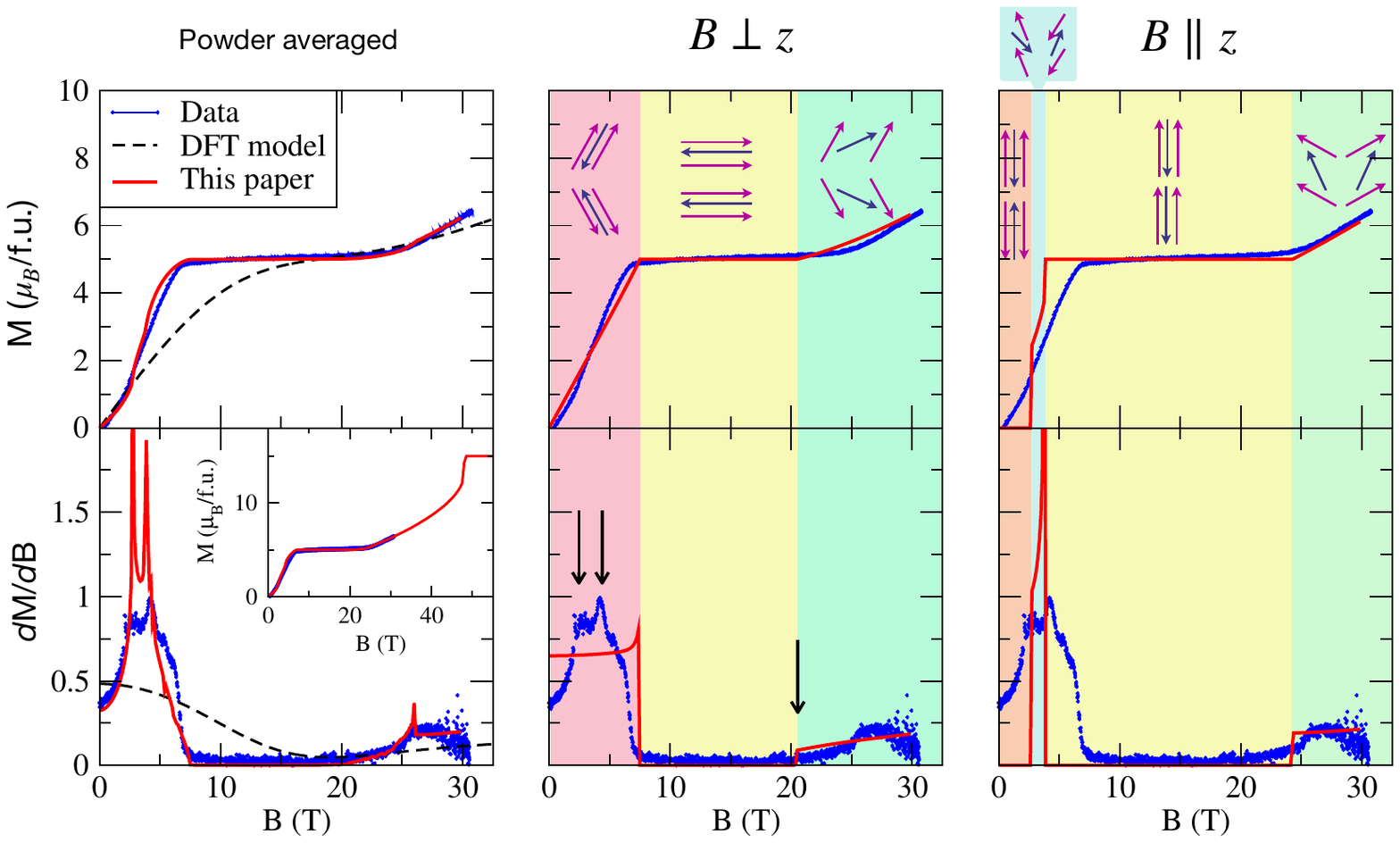}
\caption{Calculated magnetization and susceptibility in comparison to powder data from Ref.~\cite{Volkova_2016}. Calculations for the DFT model \eqref{eq_dft} from Ref.~\cite{Volkova_2016} and model \eqref{eq_model} are shown. Three panels illustrate powder averaged calculation and results for two principal field directions. Field-induced magnetic structures are shown. The arrows indicate critical fields used for fitting Hamiltonian parameters. The inset illustrates magnetization simulation in a wider region of fields, including the saturation field $H_s\approx 48$ T.}
\label{fig:mag}
\end{figure*}
In order to extract exchanges of the Hamiltonian \eqref{eq_ham}, appropriate for $\gamma$-Mn$_3$(PO$_4$)$_2$, we perform fitting of the magnetization and magnetic susceptibility data from Ref.~\cite{Volkova_2016}. This data is reproduced in Fig.~\ref{fig:mag}, where one can see several outstanding features. There is not only the magnetization plateau between 7 T and 21 T, but also two singularities around 2 T and 4 T, clearly seen in the susceptibility data.

First, one can see that the Heisenberg model with exchanges obtained by DFT \eqref{eq_dft} does not reproduce any of the aforementioned features (shown with the dashed line in Fig.~\ref{fig:mag}). For instance, additional interactions, such as biquadratic exchange and single-ion anisotropy, are necessary for stabilization of the plateau. However, as we show in the Appendix \ref{app1}, adding only one of them is also not enough.

Therefore, we explored Hamiltonian parameter phase space to match observed singularities of the magnetization data. For simplicity, we fixed the ratio of $b_i/J_i$ to be the same for $i=1,2,3$. We should note that the data from Ref.~\cite{Volkova_2016} is from the powder sample. Thus, while Heisenberg exchange and biquadratic exchange preserve SU(2) symmetry and do not exhibit grain orientation dependence, the presence of single-ion anisotropy means that calculated magnetization needs to be averaged over all possible field directions.

While, in principle, it could make fitting data extremely cumbersome, we found that observed singularities can be traced back to magnetic phase transitions for grains whose easy axis is aligned either parallel or perpendicular to the magnetic field. For instance, the transition to the plateau state for $B\parallel z$ happens at a lower field than for $B\perp z$ (see Fig.~\ref{fig:mag}), and critical field changes gradually with the field direction. Therefore, the transition for the powdered sample, after averaging, is determined only by the largest critical field - for grains with $B\perp z$. Similarly, the transition from the plateau state is determined by the grains with the lowest critical field, also for $B\perp z$. This allows us to use only calculations for particular field directions to establish boundaries on the values of exchanges. In addition, we also found that two low-field singularities of the susceptibility data originate from spin-flop-like metamagnetic transitions for grains with $B\parallel z$, and are not affected by powder averaging.

\begin{figure*}
\centering
\includegraphics[width=\linewidth]{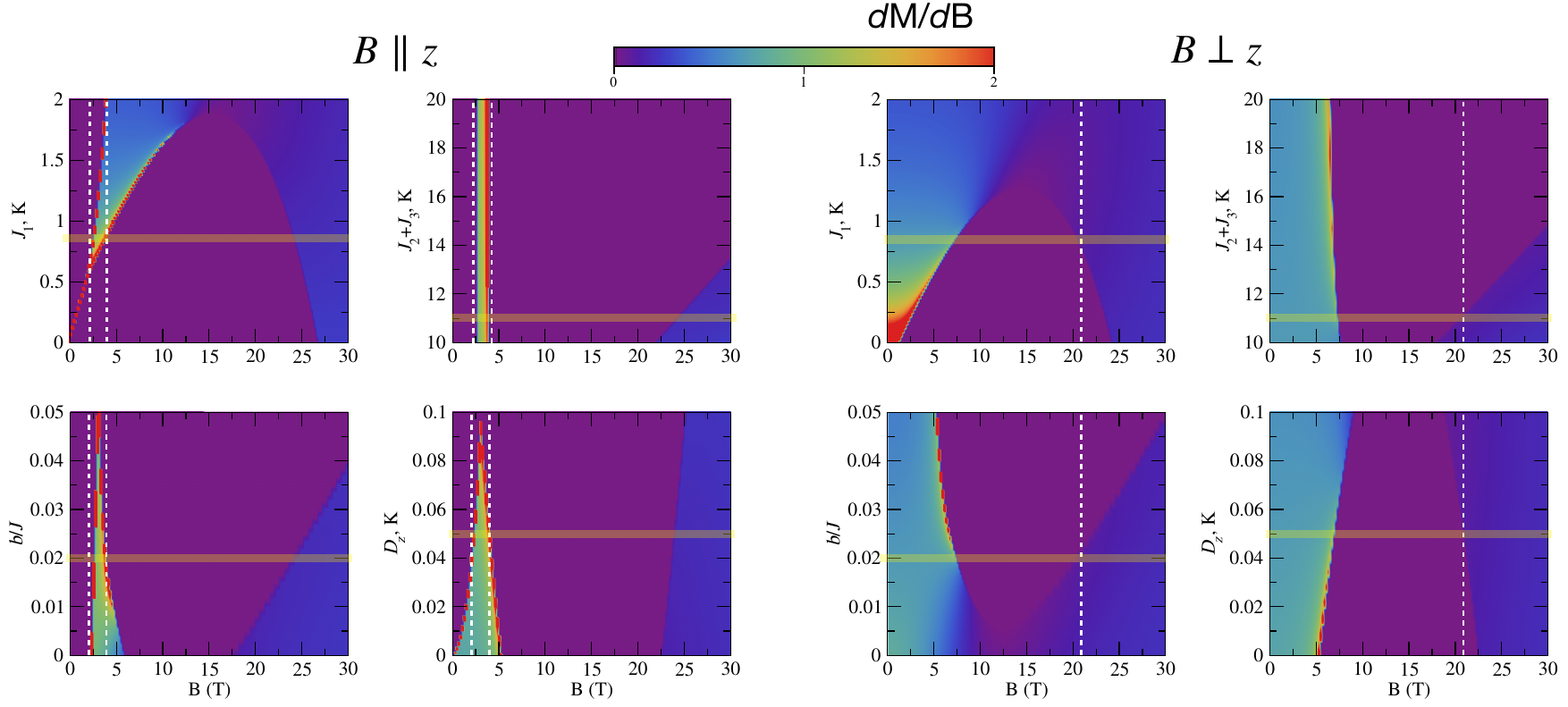}
\caption{Intensity plots of magnetic susceptibility, as a function of Hamiltonian parameters and magnetic field, for two principal directions of magnetic field. For each panel, one parameter is varied while the others are fixed at values provided in Eq.~\eqref{eq_model}. Dashed lines indicate critical fields from the susceptibility data. Parameters of model \eqref{eq_model}are shown with yellow lines.}
\label{fig:panel}
\end{figure*}

We performed a fitting procedure using the DFT model \eqref{eq_dft} as a starting point. First, as one can see in Fig.~\ref{fig:panel}, where  an intensity plot of magnetic susceptibility as a function of Hamiltonian parameters is shown, the critical field of the lower boundary of the plateau strongly depends on $J_1$ and is much less dependent on other parameters. We use it to fix $J_1=0.85\text{ K}$ to set the lower boundary at 7 T. Next, as one can see from Eq.~\eqref{eq_ham}, since positive biquadratic coupling prefers collinear states, the width of the plateau is strongly affected by the value of $b_i$. In fact, the upper boundary of the plateau mostly depends on $b/J$ and $J_2+J_3$, which we set at 21 T. Finally, as one can see in the panel for $B\parallel z$, two low-field metamagnetic transitions are strongly affected by the single-ion anisotropy $D_z$. Thus, by matching these two singularities at 2 T and 4 T, we arrive at the full model
\begin{align}
    J_1=0.85\text{ K},~J_2+J_3=11\text{ K},~b_i/J_i=0.02,~D_z=0.05\text{ K}.
    \label{eq_model}
\end{align}
The result is shown in Fig.~\ref{fig:mag} with the red line. It exhibits solid agreement with the experimental data. One issue is that calculated singularities are much sharper than the experimental data, which is most likely due to the finite temperature effects. This parameter set predicts a saturation field around 48 T, as shown in the inset of the left panel of Fig.~\ref{fig:mag}, and it can be measured in possible future experiments to verify this model.

Note that since classical energy \eqref{eq_ecl} only depends on the combinations of $J_2+J_3$ and $b_2+b_3$, this analysis of fitting magnetization data cannot provide $J_2$ and $J_3$ independently. In order to have a full model, we assume that the ratio of exchanges is the same as in the DFT calculations, thus arriving at $J_2=3.4\text{ K},~J_3=7.6\text{ K}$.

Note that the ground state of the Hamiltonian is not frustrated, and for both the DFT model \eqref{eq_dft} and the revised model \eqref{eq_model} magnetic structure consists of spins of opposing directions on Mn1 and Mn2 sites, with the anticollinear order between the layers, see also Fig.~\ref{fig:lattice}(c).

\section{Magnetic phase transitions}
\label{sec_phases}
In this section, we would like to outline field-induced states observed in our calculations. Due to the presence of single-ion anisotropy, the magnetic states are different depending on the grain orientation. We provide a description of field-induced states for two principal field directions, with the intermediate situations being not as representative.

For $B \perp z$, there are two field-induced transitions, as one can see in Fig.~\ref{fig:mag}. In zero field, the ground state antiferromagnetic configuration consists of anti-aligned Mn1 and Mn2 spins, which are pointing along the $z$ axis due to the single-ion anisotropy. When a small perpendicular field is applied, magnetic moments start canting towards the field direction, keeping the in-plane anticollinear structure. At the first critical field, this canted state gradually transforms into the 1/3 plateau. The biquadratic coupling, which favors collinear structures, stabilizes this plateau, until the second critical field is reached. The anticollinear in-plane structure is lost above the second critical field, but the $z$ components of magnetic moments are still anticollinear between the planes due to the antiferromagnetic $J_1$ interaction. The spins keep canting towards the field direction until saturation is achieved. These magnetic configurations are schematically shown in the middle panel of Fig.~\ref{fig:mag}.

For $B \parallel z$, the zero-field configuration actually survives for some values of magnetic field due to the single-ion anisotropy, until a spin-flop transition is reached. A small window of a canted state exists where the spins keep an inter-plane antiferromagnetic configuration. Increasing the field leads to the plateau state, until a canted state with spins pointing along the field direction is reached. This state gradually evolves into a fully saturated phase at high fields. These magnetic states are illustrated in the right panel of Fig.~\ref{fig:mag}.

\section{Magnetic excitations}
\label{sec_spectrum}
\begin{figure*}
\centering
\includegraphics[width=\linewidth]{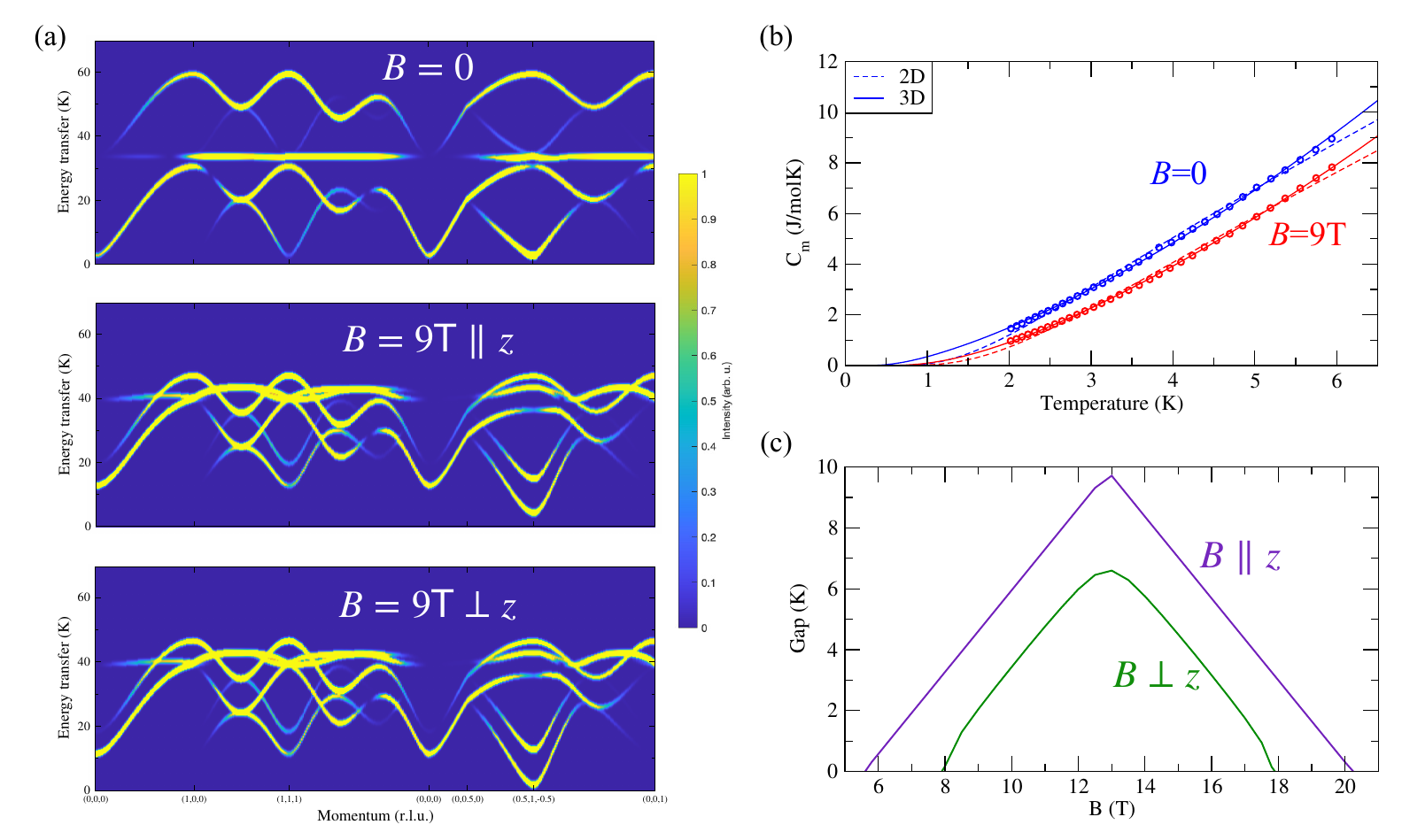}
\caption{(a) Dynamical structure factor calculated using spin-wave theory for $B=0$ and $B=9$ T, in the 1/3 plateau regime, for several high-symmetry directions in the Brillouin zone. Results for two principal field directions are shown. (b) Specific heat data for $B=0$ and $B=9$ T (blue and red circles), reproduced from Ref.~\cite{Volkova_2025}, and its low-temperature fits. Best fits are achieved by fitting with Eq.~\eqref{eq_cm_fit} with $d=3$ and values of gaps $\Delta_0=2.1\text{ K}$ and $\Delta_9=4.2\text{ K}$, for $B=0$ T and $B=9$ T respectively. (c) Gap of spin-wave excitations as a function of magnetic field (in two principal directions) in the 1/3 plateau phase.}
\label{fig:spectrum}
\end{figure*}

We also benchmarked our model against the measurements of magnetic specific heat from Ref.~\cite{Volkova_2025}. There, the value of the gap of magnetic excitations was extracted from low-temperature fits of the data. We use spin-wave theory to calculate magnetic excitations both in the zero-field antiferromagnetic state and the 1/3 ferrimagnetic plateau state. The results (obtained using SpinW package \cite{tothlake}) are shown in Fig.~\ref{fig:spectrum}(a) as an intensity plot of the dynamical structure factor
\begin{align}
\mathcal{S}(\mathbf{q},\omega)=\sum_{\alpha,\beta}\left(\delta_{\alpha \beta}-\frac{q_\alpha q_\beta}{q^2}\right)\mathcal{S}^{\alpha \beta}(\mathbf{q},\omega),
\end{align}
where the dynamical spin correlation function is given by
\begin{align}
\mathcal{S}^{\alpha \beta}(\mathbf{q},\omega)=\frac{1}{\pi} \text{Im} \int_{-\infty}^\infty dt  e^{i\omega t} \  i\langle \mathcal{T} S^\alpha_\mathbf{q}(t) S^\beta_{-\mathbf{q}}(0)\rangle.
\end{align}

In zero field, spin-wave theory yields the spectrum with a gap of $2.8\text{ K}$. In the plateau regime, for $B=9$ T and $B\parallel z$ we obtain a gap of $4.6\text{ K}$, while for $B\perp z$ the gap is $2.0\text{ K}$.

The low-temperature fits of the magnetic heat capacity data in Ref.~\cite{Volkova_2025} yielded the gap of excitations of $0.5\text{ K}$ for $B=0$ and $1.4\text{ K}$ for $B=9$ T. Here we would like to revisit fitting of that data. 

Heat capacity of the gapped spectrum $\varepsilon_\mathbf{k} \approx \Delta+A k^2$
can be approximated for low temperature $T\ll \Delta$ by \cite{fingerprinting}
\begin{align}
    C_m\approx e^{-\Delta/T}\left[ \Delta^2 T^{d/2-2}+\frac{d+2}{2}\Delta T^{d/2-1}+\frac{d+2}{2}T^{d/2}\right],
    \label{eq_cm_fit}
\end{align}
where $d$ is the dimensionality of the magnetic model. We fitted the experimental data for $B=0$ T and $B=9$ T in order to extract the gap of excitations. The results are shown in Fig.~\ref{fig:spectrum}(b), where we present calculations for $d=2$ and $d=3$. As one can see, even though one might suspect a lower dimensionality of the excitations due to $J_1 \ll J_2,J_3$, the fit with 3D excitations works much better than the one for 2D. Using Eq.~\eqref{eq_cm_fit}, we obtain for $B=0$ T
\begin{align}
    \Delta_0=2.2\text{ K}
\end{align}
which is in a solid agreement with the spin-wave theory value of $2.8\text{ K}$.  

Fitting of the $B=9\text{ T}$ data is not as straightforward due to the powder averaging of the grain directions. However, the extracted value of $\Delta_9=4.4\text{ K}$ falls within the range of gaps for two principal directions: $2.0$ K for $B\perp z$ and $4.6\text{ K}$ for $B\parallel z$. The values of the gap of spin-wave excitations for other fields in the plateau regime are shown in Fig.~\ref{fig:spectrum}(c), where a non-monotonic behavior is observed, similar to triangular-lattice antiferromagnets \cite{Kamiya_2018}.

\section{Conclusion}
\label{sec_conclusion}
While usually neutron scattering is required to determine parameters of the exchange Hamiltonian, here we showed that a peculiar sequence of field transitions can provide sufficient information. Moreover, as we showed, even the powder sample data can be deconstructed to be adequately fitted.

In particular, we showed that the 1/3 plateau and metamagnetic transitions of $\gamma$-Mn$_3$(PO$_4$)$_2$ can only be explained with the model that extends standard Heisenberg exchange with both biquadratic interaction and easy-axis anisotropy. Fitting critical fields of singularities of magnetic susceptibility data from Ref.~\cite{Volkova_2016} can be performed by using simulations of magnetic configurations for two principal field directions: parallel and perpendicular to the easy axis. As we show, powder averaging of grains with different easy-axis directions does not affect these critical fields and does not obfuscate this information.

The model for $\gamma$-Mn$_3$(PO$_4$)$_2$ that we obtain is the trimer-based decorated square lattice with weakly coupled layers. The interplay of trimers with small but nonnegligible biquadratic exchange and easy-axis anisotropy yields a unique sequence of field-induced transitions in $\gamma$-Mn$_3$(PO$_4$)$_2$, which include the 1/3 magnetization plateau. The model for $\gamma$-Mn$_3$(PO$_4$)$_2$ that we obtain, where terms beyond Heisenberg model are necessary, is in line with other Mn-based magnets \cite{Owen_1963,Edgar_1980,Hennion_2006,Pranzas_2006,Singleton_2016,Shatruk_2021}.

Furthermore, we were able to explain spectral properties of $\gamma$-Mn$_3$(PO$_4$)$_2$, such as the presence of the gap in the magnetic spectrum due to the breaking of SU(2) symmetry by easy-axis anisotropy. We performed spin-wave calculations, and the values of the gaps of excitations in zero field and the plateau regime are in solid agreement with the fits of low-temperature specific heat data from Ref.~\cite{Volkova_2025}. 
\section{Acknowledgements}
P.A.M. acknowledges support from BASIS grant 24-1-3-11-1.

\appendix

\section{Alternative models}
\label{app1}

\begin{figure*}
\centering
\includegraphics[width=\linewidth]{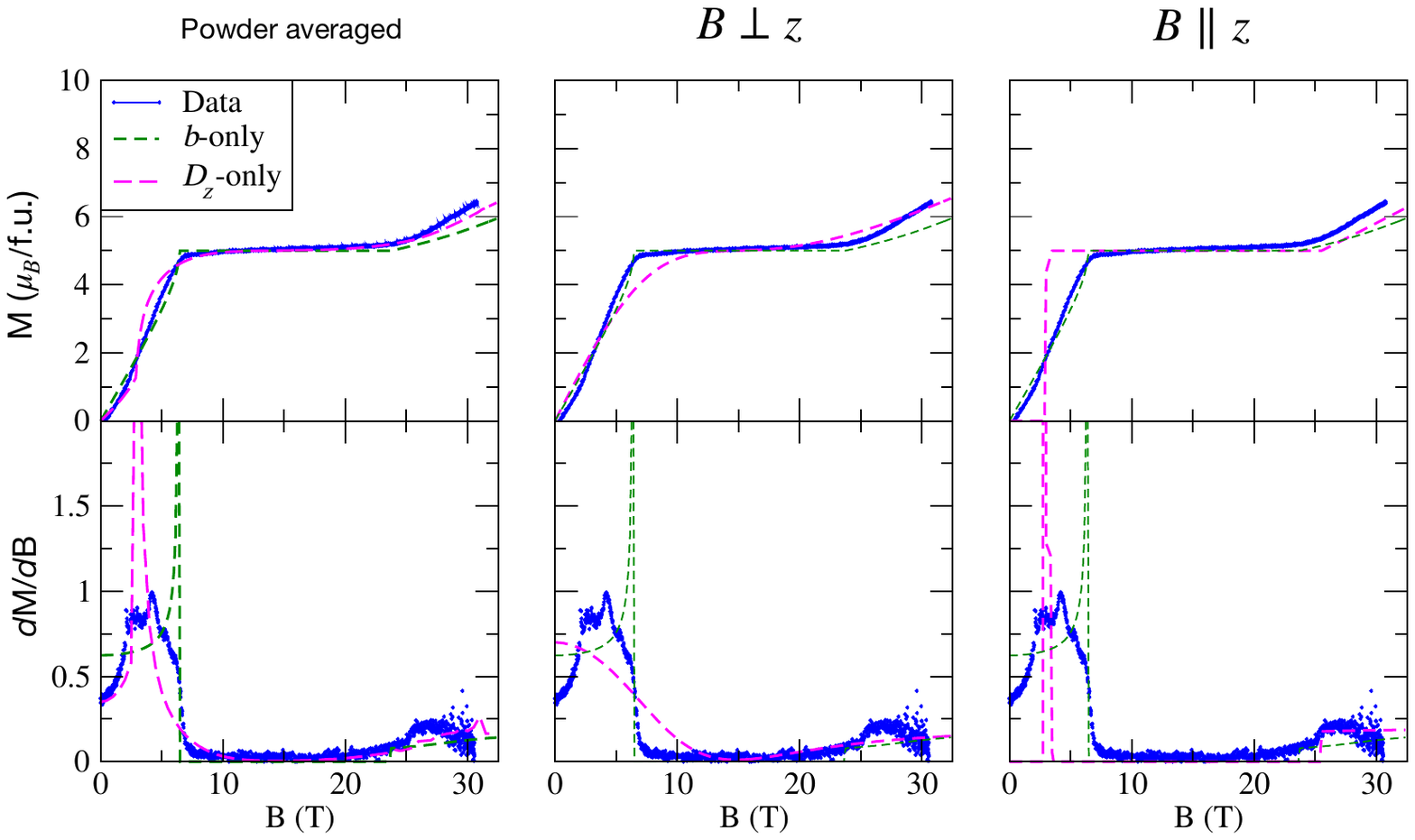}
\caption{Calculated magnetization and susceptibility in comparison to data from Ref.~\cite{Volkova_2016}. Calculations for the tuned models where, in addition to Heisenberg exchange, either biquadratic coupling or single-ion anisotropy is included. Three panels illustrate powder-averaged calculation and results for two principal field directions. One can see that neither of the models can explain observed magnetic phase transitions.}
\label{figapp:mag}
\end{figure*}

In order to show that the full exchange model for $\gamma$-Mn$_3$(PO$_4$)$_2$ requires both biquadratic exchange and easy-axis anisotropy, here we present calculations of magnetization and magnetic susceptibility where only one of these terms is included along with the Heisenberg exchange. The results are shown in Fig.~\ref{figapp:mag}. 

First, we used a model:
\begin{align}
    J_1=1.0\text{ K},~J_2+J_3=14\text{ K},~b_i/J_i=0.02,~D_z=0,
\end{align}
without single-ion anisotropy. One can see that even though the 1/3 plateau is stabilized, the behavior of magnetic susceptibility is drastically different from the experimental data.

Second, we also performed calculations for the model
\begin{align}
    J_1=0.8\text{ K},~J_2+J_3=14\text{ K},~D_z=0.12\text{ K},~b_i=0,
\end{align}
where biquadratic exchange is not included. The main discrepancy for this model is that the plateau is stabilized only for grains whose easy axis is mostly aligned with the field. Therefore, as one can see from the left panel of Fig.~\ref{figapp:mag}, there are no sharp transitions to the plateau state. Moreover, the low-field susceptibility is also not reproduced adequately.

\bibliography{mn3po4_bib}

\end{document}